\documentclass[twocolumn,aps,showpacs,nofootinbib,superscriptaddress,prd]{revtex4-2}
\usepackage{mathrsfs}
\usepackage{epsfig}
\usepackage{textcomp}
\usepackage{amsmath}
\usepackage{graphicx}
\usepackage{amssymb}
\usepackage{epstopdf}
\usepackage{slashed}
\usepackage{mathtools}
\usepackage{color}
\usepackage{slashed}
\usepackage{setspace}
\usepackage{subfigure}
\usepackage{float}
\usepackage{caption}
\usepackage{slashed}
\usepackage{bbm}
\usepackage[colorlinks,linkcolor=blue]{hyperref}
\allowdisplaybreaks[3]

\begin{document}

\title{Reduction of planar double-box diagram for single-top production via auxiliary mass flow}

\author{Najam ul Basat}
\email{najam@ihep.ac.cn}
\affiliation{Institute of High Energy Physics, Chinese Academy of Sciences, Beijing
    100049, China}
\affiliation{School of Physics Sciences, University of Chinese Academy of Sciences,
    Beijing 100039, China}

\author{Zhao Li}
\email{zhaoli@ihep.ac.cn}
\affiliation{Institute of High Energy Physics, Chinese Academy of Sciences, Beijing
    100049, China}
\affiliation{School of Physics Sciences, University of Chinese Academy of Sciences,
    Beijing 100039, China}
\affiliation{Center of High Energy Physics, Peking University, Beijing 100871, China}

\author{Yefan Wang}
\email{wangyefan@ihep.ac.cn}
\affiliation{Institute of High Energy Physics, Chinese Academy of Sciences, Beijing
    100049, China}
\affiliation{School of Physics Sciences, University of Chinese Academy of Sciences,
    Beijing 100039, China}

\begin{abstract}
The single-top production is an important process at the LHC to test the Standard Model (SM) and search for the new physics beyond the SM. Although the complete next-to-next-to-leading order (NNLO) QCD correction to the single-top production is 
crucial, this calculation is still challenging at present.  In order to efficiently reduce the NNLO single-top amplitude, we improve the auxiliary mass flow (AMF) method by introducing the $\epsilon$ truncation.
For demonstration we choose one typical planar double-box diagram for the $tW$ production. It is shown that one coefficient of the form factors on its amplitude can be systematically reduced into the linear combination of 198 scalar integrals.
\end{abstract}
\maketitle
\section{Introduction} 
The top quark is the heaviest elementary particle in the Standard Model (SM).  
In 1994, the discovery of the top quark at the Tevatron \cite{D0:1995jca,Abe:1995hr} meant the third generation of fermions in the SM is complete. Meanwhile the top quark is the only flavor that can decay before hadronization. This unique property provides an opportunity to directly measure the properties of the top quark. According to the fermion mass law \cite{Weinberg:1967tq}, the top quark mass is related to the Yukawa strength between the top quark and the Higgs boson field. After the discovery of Higgs boson at the Large Hadron Collider (LHC), this coupling can be directly studied by the ATLAS and CMS collaborations \cite{Aad:2015upn,Aad:2015eto,Aad:2015yem}. Therefore, the top quark also plays an important role in the study on the electroweak symmetry breaking (EWSB). 

At the hadron colliders, the dominant contribution to the top quark production is the top-pair production via the strong interaction, such as $q\bar{q}\to t\bar{t}$ and $gg\to t\bar{t}$. This process was discovered two decades ago \cite{D0:1995jca,Abe:1995hr}.
Then the next largest contribution to the top quark production is the single-top production via the electroweak interaction, which was observed in 2009 at the Tevatron \cite{Aaltonen:2009jj,Abazov:2009ii} for the first time. Compared to the top-pair production, the $Wtb$ vertex is included in the single-top production, and can be directly used to measure the Cabibbo–Kobayashi–Maskawa(CKM) matrix element $|V_{tb}|$ without assuming unitarity \cite{Alwall:2006bx,Cao:2015qta} and the extraction of the top quark mass \cite{Sirunyan:2017huu,Alekhin:2016jjz}. On the other hand, the single-top production can be a sensitive probe to search for the new physics beyond the SM (BSM). For instance, the single-top production could be sensitive to the new heavy gauge boson $W'$ \cite{Malkawi:1996fs,Hsieh:2010zr,Cao:2012ng}, the new fermions \cite{Cao:2006wk,Berger:2011hn,Berger:2011xk} or the new scalars \cite{Drueke:2014pla,Cao:2013ud}. And it has been found that the single-top production provides a comfortable agreement with 2HDM$+\alpha$ model \cite{Pani:2017qyd,Bauer:2017ota,Abe:2018bpo,Aad:2020zmz} to search for the Dark matter (DM).

At the LHC, there are three major modes for the single-top production : $s$-channel, $t$-channel and the $tW$ production channel. The first two channels have been observed at the Tevatron \cite{Abazov:2009ii,Aaltonen:2009jj}. And only recently the third channel, $tW$ production channel, was observed at the LHC \cite{Chatrchyan:2014tua}. 
Beside the progress of experiments, the precise theoretical predictions are demanded to match the high accuracy of experiment measurements. And the precise theoretical predictions can play vital roles in extracting
important information from the experiment data. For all above three channels, the next-to-leading order (NLO) QCD corrections have been investigated \cite{Bordes:1994ki,Smith:1996ij,Zhu:2002uj,Harris:2002md,Sullivan:2004ie,Campbell:2004ch,Cao:2004ky,Cao:2004ap,Cao:2005pq,Cao:2008af,Campbell:2009gj,Heim:2009ku}. For the next-to-next-to-leading order (NNLO) corrections, many approximate results based on soft gluon resummation have been obtained \cite{Mrenna:1997wp,Kidonakis:2006bu,Kidonakis:2007ej,Kidonakis:2010ux,Kidonakis:2010dk,Kidonakis:2011wy,Cao:2018ntd,Sun:2018byn,Li:2019dhg}. The NNLO QCD correction to $t$-channel under structure function approximation have been calculated in \cite{Brucherseifer:2014ama}. Also the NNLO calculations including top-quark leptonic decay under structure function approximation and narrow width approximation have developed in recent years \cite{Berger:2016oht,Berger:2017zof,Liu:2018gxa,Campbell:2020fhf}. And the next-to-next-to-next-to-leading order ($\text{N}^3$LO) soft-gluon corrections for the $tW$ production has been studied \cite{Kidonakis:2016sjf}.

In the calculation of the multi-loop Feynman diagrams, the amplitude generally needs to be reduced into
linear combination of the master integrals, which can be further evaluated analytically or numerically. The first key step of amplitude reduction is the tensor reduction, which is used to separate the loop momenta from fermion chains or polarization vectors. The conventional approaches to the tensor reduction include the projection method \cite{Binoth:2002xg,Glover:2004si,Wang:2019rhq} and the Tarasov's method \cite{Tarasov:1996br}. However, in some multi-scale processes, these approaches can be too complicated due to the difficulty of the inverse matrix or the dimension shift. Beside the conventional approaches, computational algebraic based
algorithms \cite{Mastrolia:2011pr,Badger:2012dp,Zhang:2012ce} and numerical unitarity method \cite{Abreu:2017xsl,Abreu:2017hqn,Badger:2017jhb,Abreu:2018zmy,Abreu:2018jgq} also have been developed in the last decade. Then after the tensor reduction, the integration by part (IBP) identities are usually implemented to reduce the scalar integrals into the master integrals. In the past decade, many algorithms and codes have been developed for the IBP reduction \cite{Laporta:2001dd,Anastasiou:2004vj,Smirnov:2008iw,Smirnov:2014hma,Smirnov:2019qkx,Maierhoefer:2017hyi,Maierhofer:2018gpa,Studerus:2009ye,vonManteuffel:2012np,Lee:2012cn,Lee:2013mka,Georgoudis:2017iza,Bendle:2019csk,Smirnov:2006tz,Lee:2008tj,Schabinger:2011dz,Larsen:2015ped,Boehm:2017wjc,vonManteuffel:2014ixa,Kosower:2018obg,Boehm:2018fpv,Chawdhry:2018awn,Mastrolia:2018uzb,Boehm:2020ijp}. 

Recently the auxiliary mass flow (AMF) method has been proposed to reduce the amplitude and the scalar integrals \cite{Liu:2018dmc,Wang:2019mnn,Guan:2019bcx}. Also it can be used to numerically evaluate the master integrals and the phase space \cite{Liu:2017jxz,Liu:2020kpc}. In amplitude reduction, this method can avoid complicated calculations of the inverse matrix and the dimension shift while the master integrals could be chosen freely.

In this paper, we introduce the truncation on $\epsilon$ to improve the efficiency of the matching procedure in the AMF method. With the help of this improvement, we can reduce one planar double-box diagram for the single-top production. Due to the complexity of multiloop multiscale diagram, we choose the integrals that include irreducible numerators to construct the set of scalar integrals. And in order to control the length of the reduction coefficients, we keep the reduction coefficients up to $\epsilon^4$, which is sufficient for the NNLO corrections. In the next section the main algorithm will be explained in detail. Then the reduction results will be shown. Finally the conclusion is made.

\section{Amplitude reduction via auxiliary mass flow}
In general the loop amplitude can be written as 
\begin{equation}
{\mathcal M} = \int
{\mathbb D}^L q
\frac{N(\{q_j\}_{j=1}^{L},\{k_e\}_{e=1}^E)}
{\prod_{i=1}^n \mathcal{D}_i^{\nu_i} },
\end{equation}
where
${\mathbb D}^L q \equiv \prod_{\ell=1}^L \left({\mathrm d}^{D}q_\ell/\left(\imath\pi^D\right)\right)$.
$\{k_e\}_{e=1}^E$ are $E$ external momenta and $\{q_j\}_{j=1}^{L}$ are $L$ loop momenta. 
$\{\mathcal{D}_i\}_{i=1}^n$ are the denominators of propagators. $N(\{q_j\}_{j=1}^{L},\{k_e\}_{e=1}^E)$ is the numerator that
may contain fermion chains or polarization vectors. In the AMF method,
all the denominators are modified as \cite{Wang:2019mnn},
\begin{equation}\label{modify}
\frac{1}{{\mathcal D}_i}\equiv \frac{1}{P_i^2-m_i^2}  \rightarrow
\frac{1}{\widetilde {\mathcal D}_i} \equiv \frac{1}{P_i^2-m_i^2+\imath\eta},
\end{equation}
where $\imath\eta$ is the auxiliary mass, $P_i\equiv Q_i+K_i$ is the momentum of the $i$-th propagator. $Q_i$ and $K_i$
are the linear combinations of respective loop momenta and external momenta.
Then we obtain the modified loop amplitude 
\begin{eqnarray}
		\widetilde {\mathcal M}(\eta) =
		\sum_{\substack{\mu_1\dots\mu_R\\ \ell_1\dots\ell_R}}
		N_{\mu_1\dots\mu_R,\ell_1\dots\ell_R}(\{k_e\}_{e=1}^E)\widetilde G^{\mu_1\dots\mu_R}_{\ell_1\dots\ell_R},
\end{eqnarray}
where
\begin{eqnarray}
	\widetilde G^{\mu_1\dots\mu_R}_{\ell_1\dots\ell_R} \equiv
	\int
	{\mathbb D}^L q~
	\frac{
		q^{\mu_1}_{\ell_1}\dots q^{\mu_R}_{\ell_R}
	}
	{
		\prod_{i=1}^n
		[(Q_i+K_i)^2-m_i^2+\imath\eta]^{\nu_i}
	}
\end{eqnarray}
is the modified tensor integral. And $N_{\mu_1\dots\mu_R, \ell_1\dots\ell_R}$ is the relevant coefficient.
At the two-loop level we can define the modified amplitude explicitly
\begin{align}
\widetilde {\mathcal M}_{uv}(\eta) &\equiv \sum_{\mu_1\dots\mu_{u+v}} N_{\mu_1\dots\mu_{u+v},\underbrace {\scriptstyle 1,\cdots,1}_u,\underbrace {\scriptstyle 2,\cdots,2}_v}(\{k_e\}_{e=1}^E)
\nonumber\\&\times
\widetilde G^{\mu_1\dots\mu_{u+v}}_{\underbrace {\scriptstyle{1,\cdots,1}}_u,\underbrace {\scriptstyle{2,\cdots,2}}_v},
\end{align}
which only include one type of the tensor integrals.
Then the two-loop modified amplitude can be written as 
\begin{eqnarray}
\widetilde {\mathcal M}(\eta) = \sum_{u,v}\widetilde {\mathcal M}_{uv}(\eta). 
\end{eqnarray}
After the Feynman parameterization \cite{Heinrich:2008si}, by using Taylor series for $\eta\to\infty$ \cite{Wang:2019mnn} we can obtain the series representation of $\widetilde {\mathcal M}_{uv}(\eta)$,
\begin{align}
\widetilde {\mathcal M}_{uv}(\eta) = \sum_i {\mathcal C}_{iuv} {\mathcal F}_i,
\end{align}
where
\begin{align}\label{Ciuv}
{\mathcal C}_{iuv} =&\eta^{\mathrm{dim}({\mathcal C}_{iuv})/2}\Bigg(\sum_{p=0}^{p_0} \sum_j\sum_{\substack{{\alpha_1,\dots,\alpha_t}\\|\alpha| = p}} \Big(a_{0p\alpha j}(D)\eta^{-p}\nonumber\\&\times s^{\alpha} I^{(vac),D}_{2,j}\Big)+\mathcal{O}(\eta^{-p_0-1})\Bigg).
\end{align}
Here ${\mathcal F}_i$ is the form factor and ${\mathcal C}_{iuv}$ is the relevant coefficient. $I^{(vac),D}_{2,j}$ represents the $j$-th 2-loop vacuum bubble master integral. 
$s \equiv (s_1,\dots,s_t)$ is the tuple of linear independent kinematic variables $\{s_1,\dots,s_t\}$. $s^{\alpha} \equiv s_1^{\alpha_1}\cdots s_t^{\alpha_t}$ is the monomial,
where $\alpha = (\alpha_1,\dots,\alpha_t)$ is a $t$-tuple of nonnegative integers. And $a_{0p\alpha j}$ is the coefficient that depends only on the space-time dimension $D$.
The explicit definitions of symbols in Eq. \eqref{Ciuv} can be found in Ref. \cite{Wang:2019mnn}.

After obtaining the series representation of ${\mathcal C}_{iuv}$, we can choose a set of integrals for the reduction. The AMF method allows one to choose integrals freely. Thus in complex multiscale process we choose the integrals that include irreducible numerators to construct the set of modified scalar integrals. Here we define two-loop modified scalar integral
\begin{align}
\widetilde I
 \equiv &\int{\mathbb D}^2 q~\frac{\Big(\prod_{e=1}^E\prod_{i=1}^2(k_e\cdot q_i)^{\rho_{ei}}\Big) \left(\prod_{l=1}^2\prod_{j=1}^l(q_j \cdot q_l)^{\sigma_{jl}}\right)}
	{\prod_{i=1}^n[(Q_i+K_i)^2-m_i^2+\imath\eta]^{\nu_i}},
\end{align}
where the exponents $\rho_{ei}$ and $\sigma_{jl}$ are nonnegative integers. And we can define a tuple $\beta =  (\beta_1,\beta_2)$, where
\begin{align}
\beta_1 &\equiv \sum_{e=1}^E\rho_{e,1}+\sigma_{1,2}+2\sigma_{1,1}
\end{align}
and
\begin{align}
\beta_2 &\equiv \sum_{e=1}^E\rho_{e,2}+\sigma_{1,2}+2\sigma_{2,2}.
\end{align}
In order to reduce $\widetilde {\mathcal M}_{uv}(\eta)$, the set of modified scalar integrals can be chosen as
\begin{align}
\left\{\widetilde I\right\}_{\beta = (u,v)}.
\end{align} 
Hence the set of modified scalar integrals has the same loop momenta rank and denominator powers with $\widetilde {\mathcal M}_{uv}(\eta)$. We use $\widetilde I_{uvk}$ to denote the $k$-th integral in $\{\widetilde I\}_{\beta = (u,v)}$. For instance, since the number of independent external momenta is 3, the reduction of $\widetilde {\mathcal M}_{2,0}(\eta)$ needs 7 modified scalar integrals,
\begin{align}
\widetilde I_{2,0,1}(\eta)\equiv&\int{\mathbb D}^2 q \dfrac{(q_1 \cdot q_1)}{\prod_{i=1}^n[(Q_i+K_i)^2-m_i^2+\imath\eta]^{\nu_i}},\nonumber\\
\widetilde I_{2,0,2}(\eta)\equiv&\int{\mathbb D}^2 q \dfrac{(k_1 \cdot q_1)^2}{\prod_{i=1}^n[(Q_i+K_i)^2-m_i^2+\imath\eta]^{\nu_i}},\nonumber\\
\widetilde I_{2,0,3}(\eta)\equiv&\int{\mathbb D}^2 q \dfrac{(k_1 \cdot q_1)(k_2 \cdot q_1)}{\prod_{i=1}^n[(Q_i+K_i)^2-m_i^2+\imath\eta]^{\nu_i}},\nonumber\\
\widetilde I_{2,0,4}(\eta)\equiv&\int{\mathbb D}^2 q \dfrac{(k_2 \cdot q_1)^2}{\prod_{i=1}^n[(Q_i+K_i)^2-m_i^2+\imath\eta]^{\nu_i}},\nonumber\\
\widetilde I_{2,0,5}(\eta)\equiv&\int{\mathbb D}^2 q\dfrac{(k_1\cdot q_1)(k_3\cdot q_1)}{\prod_{i=1}^n[(Q_i+K_i)^2-m_i^2+\imath\eta]^{\nu_i}},\nonumber\\
\widetilde I_{2,0,6}(\eta)\equiv&\int{\mathbb D}^2 q\dfrac{(k_2\cdot q_1)(k_3\cdot q_1)}{\prod_{i=1}^n[(Q_i+K_i)^2-m_i^2+\imath\eta]^{\nu_i}},\nonumber\\
\widetilde I_{2,0,7}(\eta)\equiv&\int{\mathbb D}^2 q\dfrac{(k_3\cdot q_1)^2}{\prod_{i=1}^n[(Q_i+K_i)^2-m_i^2+\imath\eta]^{\nu_i}}.
\end{align}
Then by using Taylor series for $\eta\to\infty$ we can obtain the series representation of $\widetilde I_{uvk}(\eta)$,
\begin{align}\label{Ik}
\widetilde I_{uvk} =&\eta^{\mathrm{dim}(\widetilde I_{uvk})/2}\Bigg(\sum_{p=0}^{p_0} \sum_j\sum_{\substack{{\alpha_1,\dots,\alpha_t}\\|\alpha| = p}} \Big(a_{kp\alpha j}(D)\nonumber\\&\times \eta^{-p}s^{\alpha} I^{(vac),D}_{2,j}\Big)+\mathcal{O}(\eta^{-p_0-1})\Bigg).
\end{align}

Analogue to the procedure in \cite{Wang:2019mnn}, by matching the form factor coefficient ${\mathcal C}_{iuv}$ and the set of modified scalar integrals $\{\widetilde I_{uvk}\}$ in the series representation, one can generate the coefficient matrix $\mathbb{M}(D)$, where $D = 4 - 2\epsilon$ is the space-time dimension. Finally the reduction problem can be transformed into the null space problem of $\mathbb{M}(D)$,
\begin{equation}\label{e1}
\mathbb{M}(D)\cdot \mathbb{X}(D)=0,
\end{equation}
where the coefficient matrix $\mathbb{M}(D)$ and the null space $\mathbb{X}(D)$ only depend on $D$. For the NNLO correction, the coefficient matrix usually can be large and complicated. Consequently the null space could be difficult to obtain, and the reduction coefficients can be very long.

To efficiently solve the null space and control the length of reduction coefficients in the dimension regularization, first we expand the coefficient matrix $\mathbb{M}(\epsilon)$ at $\epsilon\to 0$,
\begin{equation}\label{e2}
\mathbb{M}(\epsilon) =  \mathbb{M}_0 + \mathbb{M}_1 \epsilon + \cdots +\mathbb{M}_m \epsilon^m+\mathcal{O}(\epsilon^{m+1}),
\end{equation}
where $\mathbb{M}_0,\cdots, \mathbb{M}_n$ are constant matrices.
Similarly the unknown null space $\mathbb{X}(\epsilon)$ can also be expanded as
\begin{equation}\label{e3}
\mathbb X(\epsilon) =  \mathbb X_0 + \mathbb X_1 \epsilon +\cdots+ \mathbb X_m \epsilon^m +\mathcal{O}(\epsilon^{m+1}).
\end{equation}
Then by substituting Eq. \eqref{e2} and Eq. \eqref{e3} 
into Eq. \eqref{e1} we can obtain a linear system of equations
\begin{align}\label{eqs}
\mathbb{M}_0 &\cdot \mathbb X_0 = 0,\nonumber\\
\mathbb{M}_0 &\cdot \mathbb X_1 + \mathbb{M}_1 \cdot \mathbb X_0  = 0,\nonumber\\
\mathbb{M}_0 &\cdot \mathbb X_2 + \mathbb{M}_1 \cdot \mathbb X_1 + \mathbb{M}_2 \cdot \mathbb X_0 = 0,\nonumber\\
&\cdots\nonumber\\
\mathbb{M}_0 &\cdot \mathbb X_m + \mathbb{M}_1 \cdot \mathbb X_{m-1} + \cdots + \mathbb{M}_m \cdot \mathbb X_0 =0.
\end{align}
Starting from the $\epsilon^0$ order, we assume that the equation 
\begin{align}
\mathbb{M}_0 &\cdot \mathbb X_0 = 0
\end{align}
have $r_0$ solutions. Then we have
\begin{align}\label{eps0}
\mathbb{M}_0 &\cdot \mathbb X_0^{(r_0)} = 0,
\end{align}
where
\begin{align}
\mathbb X_0^{(r_0)} \equiv (\mathbb X_0^1,\mathbb X_0^2,\cdots,\mathbb X_0^{r_0}).
\end{align}
Since the linear combinations of $\{\mathbb X_0^1,\mathbb X_0^2,\cdots,\mathbb X_0^{r_0}\}$ are also the solutions of Eq. \eqref{eps0}, the null space equation at $\epsilon^1$ order becomes
\begin{align}\label{eps1_2}
\left(\mathbb{M}_1 \cdot \mathbb X_0^{(r_0)},\mathbb{M}_0\right)
\begin{pmatrix}
\mathbb C_0^{(r_0,r_1)}\\ \mathbb X_1^{(r_1)}
\end{pmatrix}
=0,
\end{align}
where
\begin{align}
\mathbb X_1^{(r_1)} \equiv (\mathbb X_1^1,\mathbb X_1^2,\cdots,\mathbb X_1^{r_1})
\end{align}
are $r_1$ solutions, and $\mathbb C_0^{(r_0,r_1)}$ is $r_0\times r_1$ constant matrix. Now up to $\epsilon^1$, the solutions can be expressed as 
\begin{align}
\mathbb X_0^{(r_1)} + \mathbb X_1^{(r_1)} \epsilon,
\end{align}
where $\mathbb X_0^{(r_1)} \equiv \mathbb X_0^{(r_0)}\cdot \mathbb C_0^{(r_0,r_1)}$.

Suppose that up to $\epsilon^p$ we have the solutions
\begin{align}
\mathbb X_0^{(r_p)} + \mathbb X_1^{(r_p)} \epsilon + \cdots +\mathbb X_p^{(r_p)}\epsilon^p.
\end{align}
Then at $\epsilon^{p+1}$ we can have
\begin{align}
\left(\mathbb{M}_{p+1} \cdot \mathbb X_0^{(r_p)}+\cdots+\mathbb{M}_{1}\cdot \mathbb X_p^{(r_p)},\mathbb{M}_0\right)\cdot
\begin{pmatrix}
\mathbb C_p^{(r_p,r_{p+1})}\\ \mathbb X_{p+1}^{(r_{p+1})} 
\end{pmatrix}
= 0.
\end{align}
Then up to the next order $\epsilon^{p+1}$ we can obtain the solution,
\begin{align}
\mathbb X_0^{(r_{p+1})} + \mathbb X_1^{(r_{p+1})}\epsilon +\cdots+\mathbb X_{p+1}^{(r_{p+1})} \epsilon^{p+1},
\end{align}
where 
\begin{align}
\mathbb X_0^{(r_{p+1})} &\equiv \mathbb X_0^{(r_p)}\cdot \mathbb C_p^{(r_p,r_{p+1})},\nonumber\\
&\cdots\nonumber\\
\mathbb X_p^{(r_p+1)} &\equiv \mathbb X_p^{(r_p)}\cdot \mathbb C_p^{(r_p,r_{p+1})}.
\end{align}
Therefore, by the iteration relations we can obtain the approximate solutions $\mathbb X^{approx}(\epsilon)$,
\begin{align}
\mathbb X^{approx}(\epsilon) =  \mathbb X_0^{(r_m)} + \mathbb X_1^{(r_m)} \epsilon + \cdots + \mathbb X_m^{(r_m)} \epsilon^m.
\end{align}

Since we take the truncation on $\epsilon$ in $\mathbb{M}(\epsilon)$ and $\mathbb{X}(\epsilon)$, the equations in Eq. \eqref{eqs} are the parts of the complete linear system of equations in Eq. \eqref{e1}. Meanwhile the approximation of the true solutions $\mathbb{X}(\epsilon)$ must exist within $\mathbb X^{approx}(\epsilon)$. And there could be some redundant solutions that satisfy
Eq. \eqref{eqs} but not satisfy Eq. \eqref{e1}. To check if there are redundant solutions in $\mathbb X^{approx}(\epsilon)$, we can observe the number of the linear independent solutions in $\mathbb X^{approx}(\epsilon)$. If it is equal to the nullity of $\mathbb{M}(\epsilon)$, which can be obtained by randomly assigning $\epsilon$ as some constant numbers, it means that there is no redundant solution. Therefore, the $\mathbb X^{approx}(\epsilon)$ is the approximation of the $\mathbb{X}$,
\begin{align}
\mathbb X^{approx}(\epsilon) = \mathbb{X}(\epsilon)|_{\epsilon^{m+1}=0}.
\end{align}
If there are redundant solutions in $\mathbb X^{approx}(\epsilon)$, 
we can use higher $\epsilon$ order expansions to introduce more constraints to the linear system in Eq. \eqref{eqs}.
Then we repeat the above procedure until there are no redundant solutions in $\mathbb X^{approx}(\epsilon)$.

Also the finite fields are implemented to improve the efficiency in null space calculations of the constant matrices. And the Chinese remainder theorem (CRT) is used to reconstruct the rational numbers. Finally we keep the reduction coefficients up to $\epsilon^{4}$ since at the two-loop level the maximum divergence of integrals is $\epsilon^{-4}$. Consequently the null space problem can be efficiently solved. And the length of reduction coefficients can be effectively controlled.

After the reduction, $\widetilde {\mathcal M}_{uv}(\eta)$ can be reduced to several modified scalar integrals
\begin{equation}
\widetilde {\mathcal M}_{uv}(\eta) =  \sum_i \sum_{k} C_{iuvk}\widetilde I_{uvk}(\eta){\mathcal F}_i,
\end{equation}
where $C_{iuvk}$ is the reduction coefficient of relevant $\widetilde I_{uvk}(\eta)$ and ${\mathcal F}_i$ for $\widetilde {\mathcal M}_{uv}(\eta)$. 
Since the set of modified scalar integrals $\{\widetilde I\}_{\beta = (u,v)}$ and $\widetilde {\mathcal M}_{uv}(\eta)$ have same denominator powers and loop momenta rank, the reduction coefficients $\{C_{iuvk}\}$ 
only depend on the numerators of amplitude and modified scalar integrals. And the auxiliary mass $\imath \eta$ only exist in the denominators of amplitude and scalar integrals.
Consequently, the reduction coefficients $\{C_{iuvk}\}$ are independent of $\eta$.

For given form factor ${\mathcal F}_i$ of amplitude, the set $\{\widetilde I_{uvk}\}$ and $\{C_{iuvk}\}$
can be ordered using certain well order relation, e.g. lexicographical ordering, for $(u,v,k)$, respectively. And $\widetilde  I_{p}$ and $C_{ip}$ can be denoted as the $p$-th element 
in the corresponding set. Finally the modified amplitude can be written as
\begin{equation}
\widetilde {\mathcal M}(\eta) = \sum_{u,v}\widetilde {\mathcal M}_{uv}(\eta) = \sum_i \sum_{p} C_{ip} \widetilde I_{p}(\eta){\mathcal F}_i
\end{equation}
Since the $C_{ip}$ is independent of $\eta$, When $\eta\to 0$, the original amplitude can be written as
\begin{align}
{\mathcal M} = \sum_i \sum_{p} C_{ip} I_{p}{\mathcal F}_i.
\end{align}
These scalar integrals $\{I_p\}$ can be further reduced into final master integrals via 
auxiliary mass flow or other methods.

\section{Planar double-box diagram in the single-top production}
\begin{figure}[H]
	\centering
	\includegraphics[width=0.35\textwidth]{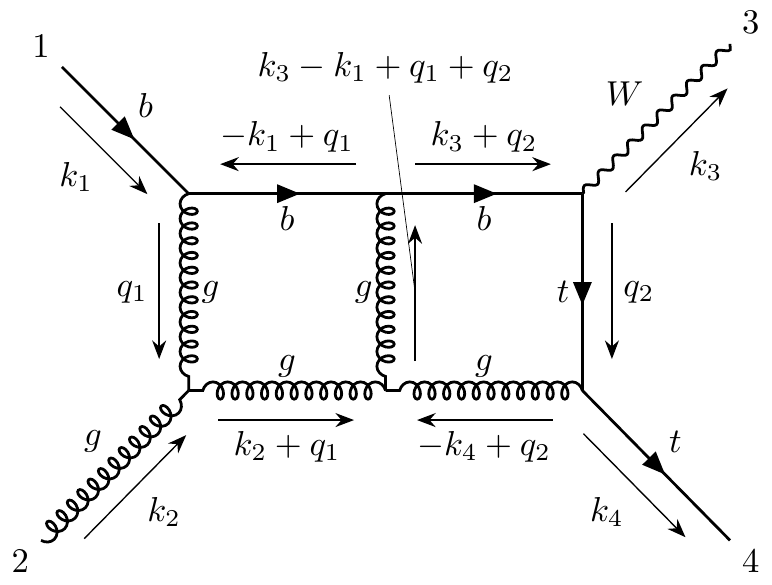}
	\caption{Double-box diagram for the $tW$ production}
	\label{draw_diagram182}
\end{figure}

In this section, we implement our improved approach on one double-box diagram of the $tW$ production process $b(k_1)+g(k_2)\rightarrow W(k_3)+t(k_4)$. The Feynman diagram is shown in Fig.\ref{draw_diagram182}, which is plotted by LaTeX package \textsc{TikZ-Feynman} \cite{Ellis:2016jkw}.
Its relevant modified amplitude can be written as
\begin{equation}
\widetilde {\mathcal M}(\eta) = \int {\mathbb D}^2 q
\frac{N(q_1,q_2,k_1,k_2,k_3,k_4)}{
\widetilde {\mathcal D}_1
\widetilde {\mathcal D}_2
\widetilde {\mathcal D}_3
\widetilde {\mathcal D}_4
\widetilde {\mathcal D}_5
\widetilde {\mathcal D}_6
\widetilde {\mathcal D}_7
},
\end{equation}
where the denominators are
\begin{align}
\widetilde {\mathcal D}_1 &= (q_1-k_1)^2 +\imath \eta,\nonumber\\
\widetilde {\mathcal D}_2 &= (q_1+k_2)^2 +\imath \eta,\nonumber\\
\widetilde {\mathcal D}_3 &= (k_3-k_1+q_1+q_2)^2 +\imath \eta,\nonumber\\
\widetilde {\mathcal D}_4 &= (k_3+q_2)^2 +\imath \eta,\nonumber\\
\widetilde {\mathcal D}_5 &=(q_2-k_4)^2 +\imath \eta,\nonumber\\
\widetilde {\mathcal D}_6 &= q_1^2 +\imath \eta,\nonumber\\
\widetilde {\mathcal D}_7 &= q_2^2-m_t^2+\imath \eta.
\end{align}
Here $m_W$ is the mass of $W$ boson and $m_t$ is the mass of top quark. And we define 
\begin{align}
s_1 &\equiv 2(k_1 \cdot k_2),\nonumber\\
s_2 &\equiv m_W^2 - 2(k_1 \cdot k_3).
\end{align}
For reader's convenience we explicitly show the numerator of this amplitude
\begin{align}
&N(q_1,q_2,k_1,k_2,k_3,k_4) = -\imath g_{49}g_{10}^2g_{11}^3 \varepsilon^{\mu_5}(k_2) \nonumber\\
\times&\bar u(k_4) \gamma^{\mu_1}(\slashed{q_2}+m_t)\slashed{\varepsilon^*}(k_3)P_L(\slashed{q_2}+\slashed{k_3})\gamma^{\mu_2}(\slashed{k_1}-\slashed{q_1})
\gamma^{\mu_3} u(k_1)\nonumber\\
\times&\{(q_1-k_2)^{\mu_4}g^{\mu_3\mu_5}+(2k_2+q_1)^{\mu_3}g^{\mu_4\mu_5}-(k_2+2q_1)^{\mu_5}g^{\mu_3\mu_4}\}\nonumber\\
\times&\{(2k_4-k_2-q_1+2q_2)^{\mu_4}g^{\mu_1\mu_2}+(2q_1+q_2+2k_2-k_4)^{\mu_1}\nonumber\\
&g^{\mu_2\mu_4}+(-q_1+q_2-k_2-k_4)^{\mu_2}g^{\mu_1\mu_4}\}.
\end{align}
From this amplitude we can extract 10 linear independent form factors,
\begin{align}
{\mathcal F}_{1} &=\bar{u}(k_4)P_L\slashed{\varepsilon}(k_2)\slashed{\varepsilon}^{*}(k_3)u(k_1),\nonumber\\
{\mathcal F}_{2} &=\bar{u}(k_4)P_L\slashed{\varepsilon}(k_2)\slashed{k_2}u(k_1)(k_2 \cdot \varepsilon^{*}(k_3)),\nonumber\\
{\mathcal F}_{3} &=\bar{u}(k_4)P_L\slashed{\varepsilon}^*(k_3)\slashed{k_2}u(k_1)(k_3 \cdot \varepsilon(k_2)),\nonumber\\
{\mathcal F}_{4} &=\bar{u}(k_4)P_L u(k_1)(k_3 \cdot \varepsilon(k_2))(k_2 \cdot \varepsilon^{*}(k_3)),\nonumber\\
{\mathcal F}_{5} &=\bar{u}(k_4)P_L u(k_1)(\varepsilon(k_2)\cdot\varepsilon^{*}(k_3)),\nonumber\\
{\mathcal F}_{6} &=\bar{u}(k_4)P_R\slashed{\varepsilon}(k_2)\slashed{\varepsilon}^{*}(k_3)\slashed{k_2}u(k_1),\nonumber\\
{\mathcal F}_{7} &=\bar{u}(k_4)P_R\slashed{\varepsilon}(k_2)u(k_1)(k_2 \cdot \varepsilon^{*}(k_3)),\nonumber\\
{\mathcal F}_{8} &=\bar{u}(k_4)P_R\slashed{\varepsilon}^*(k_3)u(k_1)(k_3 \cdot \varepsilon(k_2)),\nonumber\\
{\mathcal F}_{9} &=\bar{u}(k_4)P_R\slashed{k_2}u(k_1)(k_2\cdot\varepsilon^{*}(k_3))(k_3\cdot\varepsilon(k_2)),\nonumber\\
{\mathcal F}_{10} &=\bar{u}(k_4)P_R\slashed{k_2}u(k_1)(\varepsilon(k_2)\cdot\varepsilon^{*}(k_3)).
\end{align} 
In projection method, to finish the tensor reduction of the amplitude, one needs to calculate the analytical inversion of the projection matrix, which includes 4 independent 
kinematic variables and space-time dimension $D$. Since the symbolic matrix is big and complex, the analytical calculation of its inversion is quite difficult.

As mentioned in last section, the modified amplitude can be decomposed into 15 parts
\begin{align}
\widetilde {\mathcal M}(\eta) =& \widetilde {\mathcal M}_{0,0}(\eta)+\widetilde {\mathcal M}_{0,1}(\eta)+\widetilde {\mathcal M}_{1,0}(\eta)+\widetilde {\mathcal M}_{0,2}(\eta)\nonumber\\&
+ \widetilde {\mathcal M}_{1,1}(\eta)+ \widetilde {\mathcal M}_{2,0}(\eta)+\widetilde {\mathcal M}_{0,3}(\eta)+\widetilde {\mathcal M}_{1,2}(\eta)\nonumber\\&
+\widetilde {\mathcal M}_{2,1}(\eta)+\widetilde {\mathcal M}_{3,0}(\eta)
+\widetilde {\mathcal M}_{1,3}(\eta)+\widetilde {\mathcal M}_{2,2}(\eta)\nonumber\\&
+\widetilde {\mathcal M}_{3,1}(\eta)+\widetilde {\mathcal M}_{2,3}(\eta)+\widetilde {\mathcal M}_{3,2}(\eta).
\end{align}
In this paper we show the reduction results for the coefficient of ${\mathcal F}_{1}$. The reductions of the other form factors can be finished in the same way. And the reduction difficulty of the other form factors is in the same level of ${\mathcal F}_{1}$.
For simplicity, we only show the scalar integrals $I_{p}$ with nonzero reduction coefficients. 
Consequently 198 out of 486 scalar integrals are remaining.
The reduction coefficients are kept up to $\epsilon^{4}$. And the length of reduction coefficients can be effectively controlled. For convenience the constant factor $\imath g_{49}g_{10}^2g_{11}^3$
is factorized out in the results. For instance,
\begin{align}
I_{198 }\equiv&\int {\mathbb D}^2 q\dfrac{(k_3\cdot q_1) (q_1\cdot q_1) (q_2\cdot q_2)}{{\mathcal D}_1{\mathcal D}_2{\mathcal D}_3{\mathcal D}_4{\mathcal D}_5{\mathcal D}_6{\mathcal D}_7}.
\end{align}
By using the in-house package \textsc{SeRA.jl}, the corresponding reduction coefficient is
\begin{align}
C_{1,198}=&\frac{2s_1m_t\left(1 +5 \epsilon +12 \epsilon^2 +24 \epsilon^3 +48 \epsilon^4\right)}{m_t^2\left(m_W^2-s_2\right)+s_2
\left(-m_W^2+s_1+s_2\right)}.
\end{align}
The complete expressions of the set of scalar integrals and their reduction coefficients can be downloaded in \url{https://github.com/zhaoli-IHEP/gbtw_Reduction_Data}. To cross check the coefficients $\{C_{1,p}\}$, we use the Tarasov's method \cite{Tarasov:1996br} and IBP reduction to reduce the original amplitude numerically. Then we also apply the numerical IBP reduction to $\{I_{p}\}$. 
Finally the two reduction results are consistent. In IBP reduction procedure we use packages \textsc{FIRE} \cite{Smirnov:2014hma,Smirnov:2019qkx} and \textsc{LiteRed} \cite{Lee:2013mka}.

\section{Conclusion}
In this paper, we improve the AMF method by taking the truncation on $\epsilon$ in the matching procedure. And we reduce one planar double-box diagram for the $tW$ production as the demonstration. The amplitude can be easily reduced into 10 form factors. One coefficient of the form factors can be easily reduced into 198 scalar integrals which include irreducible numerators. And the length of the reduction coefficients can be effectively controlled. This approach can be implemented on some other important processes in the future, such as the other diagrams in the NNLO $tW$ production.

\section{Acknowledgments}
This work was supported by the National Natural Science Foundation of China under Grant No. 11675185 and 12075251. Najam ul Basat would like to acknowledge financial support from CAS-TWAS President's Fellowship Program 2017. The authors want to thank Yan-Qing Ma, Jian Wang and Yang Zhang for helpful discussions.

	\bibliography{single_top}
\end{document}